\documentclass[graybox]{svmult}

\usepackage[T1]{fontenc}

\usepackage{type1cm}        
                            
\usepackage{makeidx}         
\usepackage{graphicx}        
\usepackage{multicol}        
\usepackage[bottom]{footmisc}

\usepackage{newtxtext}       %
\usepackage[varvw]{newtxmath}       

\usepackage{svg}
\usepackage{float}
\usepackage{array}
\usepackage{adjustbox}
\usepackage{caption}
\usepackage{booktabs}
\usepackage{color, colortbl}
\usepackage{xcolor, soul}

\sethlcolor{yellow}

\usepackage[bookmarks=false]{hyperref}
\hypersetup{
    colorlinks      = {true},
    linkcolor       = {blue}, 
    linkbordercolor = {white},
    citecolor       = {blue},
    urlcolor        = {blue}
}
      
\usepackage[super]{nth}

\usepackage{comment}
\usepackage{todonotes}

\usepackage{doi}
\usepackage{orcidlink}

\makeindex             

\begin{document}

\title{Wireless Crowd Detection for Smart Overtourism Mitigation}

\author{
Tomás Mestre Santos \inst{{1} \orcidlink{0009-0004-0255-8638}} \and
Rui Neto Marinheiro \inst{{1} \orcidlink{0000-0002-0385-8876}} \and 
Fernando Brito e Abreu \inst{{1} \orcidlink{0000-0002-9086-4122}}
}

\institute{
    1) Instituto Universitário de Lisboa (ISCTE-IUL), Lisboa, Portugal,\\
    \email{{tmmss1, rui.marinheiro, fba}@iscte-iul.pt}
}

\maketitle

\abstract{
Overtourism occurs when the number of tourists exceeds the carrying capacity of a destination, leading to negative impacts on the environment, culture, and quality of life for residents. By monitoring overtourism, destination managers can identify areas of concern and implement measures to mitigate the negative impacts of tourism while promoting smarter tourism practices. This can help ensure that tourism benefits both visitors and residents while preserving the natural and cultural resources that make these destinations so appealing.
\\
This chapter describes a low-cost approach to monitoring overtourism based on mobile devices’ wireless activity. A flexible architecture was designed for a smart tourism toolkit to be used by Small and Medium-sized Enterprises (SMEs) in crowding management solutions, to build better tourism services, improve efficiency and sustainability, and reduce the overwhelming feeling of pressure in critical hotspots.
\\
The crowding sensors count the number of surrounding mobile devices, by detecting trace elements of wireless technologies, mitigating the effect of MAC address randomization. They run detection programs for several technologies, and fingerprinting analysis results are only stored locally in an anonymized database, without infringing privacy rights. After that edge computing, sensors communicate the crowding information to a cloud server, by using a variety of uplink techniques to mitigate local connectivity limitations, something that has been often disregarded in alternative approaches.
\\
Field validation of sensors has been performed on Iscte’s campus. Preliminary results show that these sensors can be deployed in multiple scenarios and provide a diversity of spatio-temporal crowding data that can scaffold tourism overcrowding management strategies.
}


\section{Introduction}
\label{sec:introduction}
\begin{figure}[!htb]
  \centering
  \includegraphics[scale=0.15]{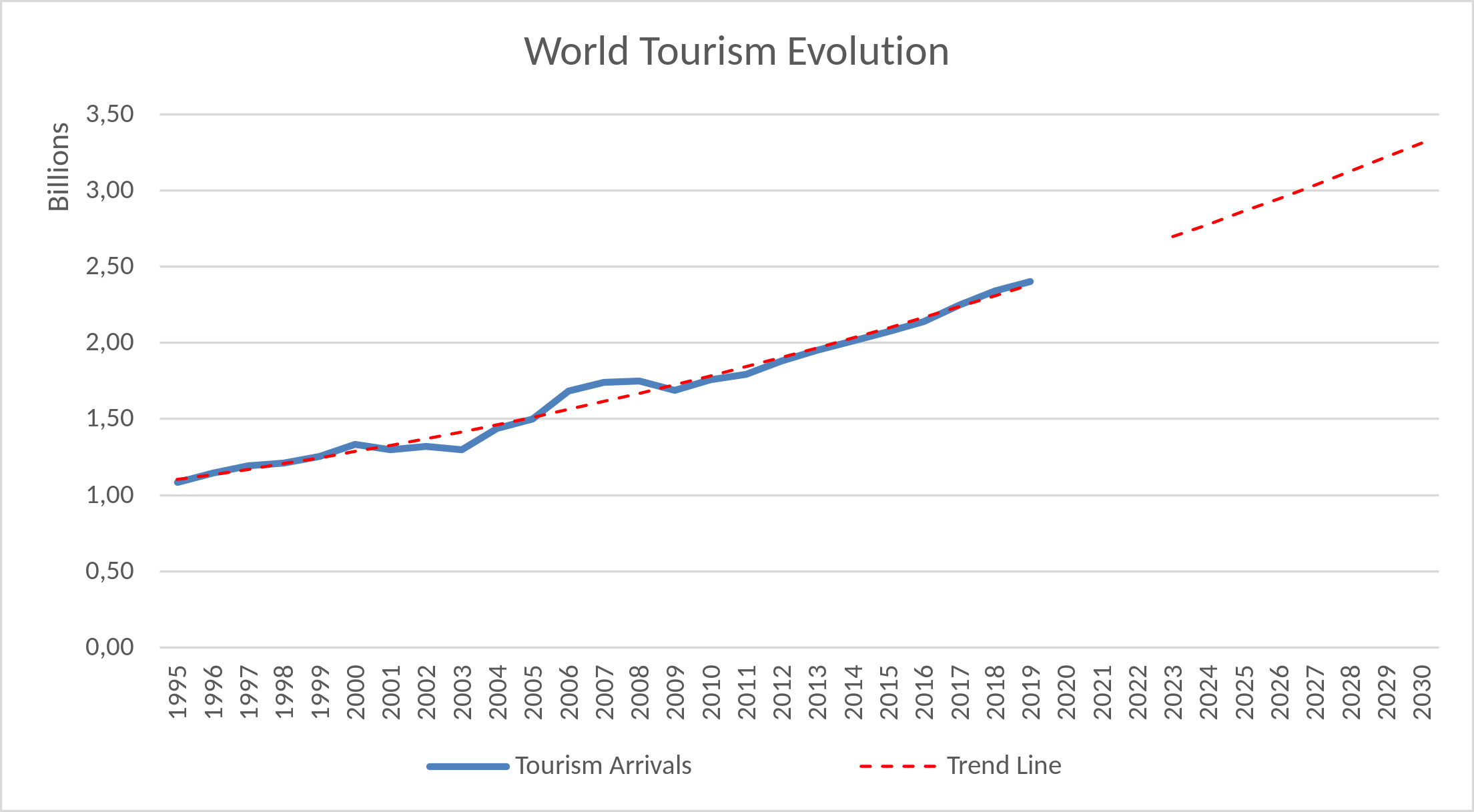}
  \caption{Worldwide evolution of tourism arrivals, based on the World Bank's data}
  \label{fig:WorldBank_TourismEvolution}
\end{figure}

The tourism sector has been growing steadily. If the pre-pandemic trend is achieved from 2023 onward, it will reach 3 billion arrivals by 2027, based on the \href{https://databank.worldbank.org/source/world-development-indicators/Series/ST.INT.ARVL}{World Bank development indicators} (see Figure \ref{fig:WorldBank_TourismEvolution}).

As a consequence, the impact of tourist activities in popular destinations has risen significantly over the years, often fostered by the proliferation of cheaper local accommodation \cite{Guttentag2015}. That increase led to exceed of carrying capacity in those destinations, a phenomenon called \emph{tourism overcrowding}, or simply \emph{overtourism}. The latter degrades visitors' quality of experience, reducing their feeling of safety, making it difficult to move around, enjoy the attractions, and use basic services, such as transportation and restoration, due to long wait times, while reducing the authenticity from the perspective of tourists \cite{tokarchuk2022}.
Overtourism also deteriorates the lives of local residents, due to an increase in urban noise, less effective urban cleaning, higher prices for basic goods and services (as businesses seek to capitalize on the increased demand), displacement caused by local accommodation, and cultural clashes when visitors fail to respect local customs, traditions, and privacy, sometimes leading to the former expressing negatively against the latter \cite{biendicho2022}.
Last, but not least, the environmental sustainability, structures, and cultural heritage of overcrowded destinations are also jeopardized, leading to a loss of authenticity \cite{Seraphin2018}. Mitigating overtourism benefits all stakeholders:

\begin{itemize}
    \item Local residents reduce their stress from over-occupation of personal space and privacy, and improve their attitude towards tourists and tourism professionals;
    \item Tourism operators speed up service delivery and quality of service;
    \item Tourists increase their visit satisfaction, with fewer delays, increased safety, and cleanliness;
    \item Local authorities improve services by making just-in-time decisions and planning more effectively urban cleaning and public safety routines, as well as reducing operating costs;
    \item Heritage managers can prevent heritage degradation more effectively, thus retaining the authenticity of destinations;
    \item Local businesses increase their share of tourism income.
\end{itemize}

Overtourism mitigation actions, such as promoting the visitation to less occupied but equally attractive areas, can be applied in recreational, cultural, or religious spots, both in indoor scenarios like palaces, museums, monasteries, or cathedrals, or in outdoor ones such as public parks, camping parks, concerts, fireworks, or video mapping shows. 

Besides assuring a better visiting experience, those actions are also necessary for security reasons (e.g., to prevent works exhibited in a museum from deteriorating or even being vandalized by exceeding room capacity), health reasons (e.g., preventing infection in pandemic scenarios by not exceeding the maximum people density specified by health authorities), or even for resource management (e.g. to reduce the intervention of security and cleaning teams).

\begin{figure}[!htb]
  \centering
  \includegraphics[scale=0.36]{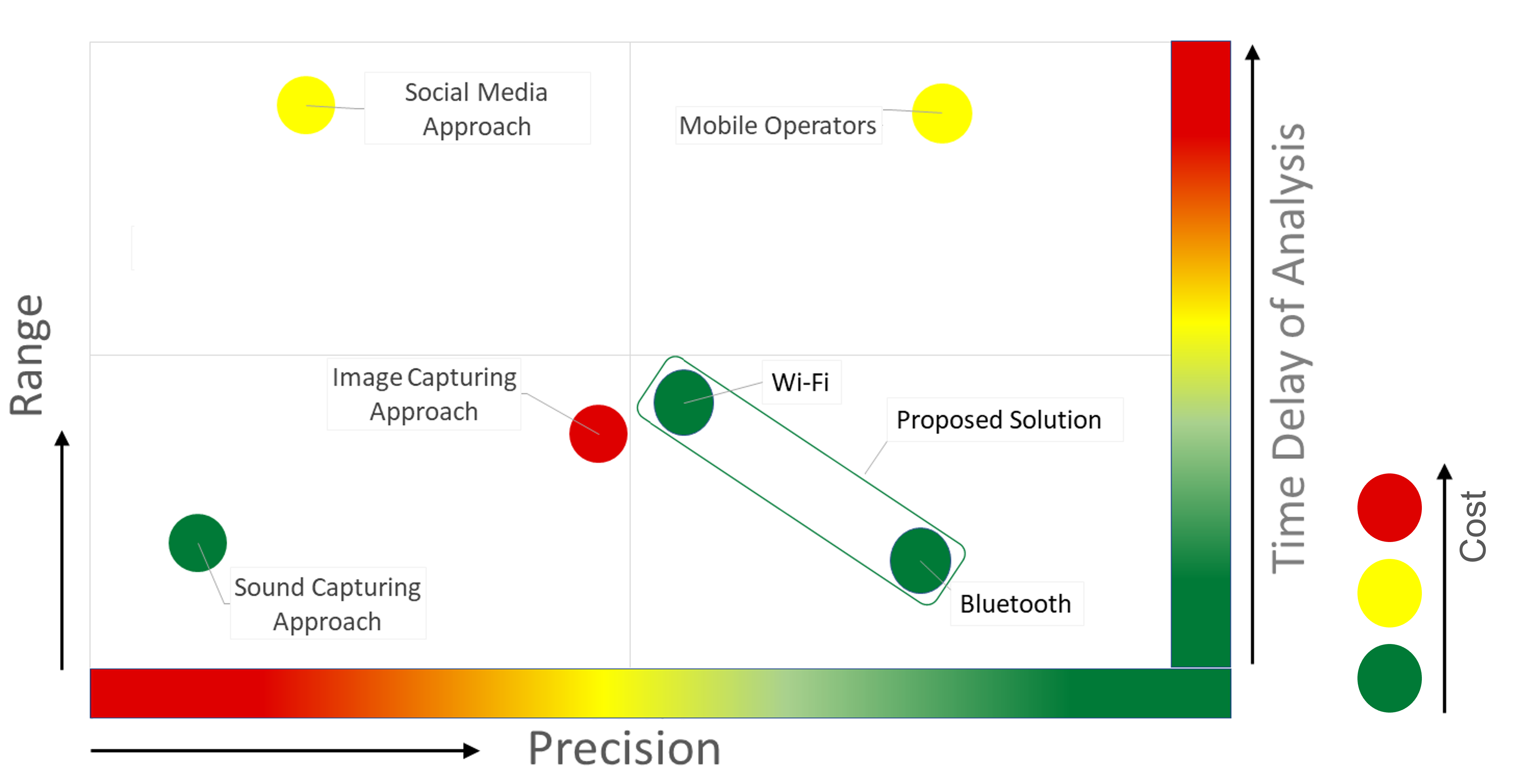}
  \caption{Different approaches for crowd counting in terms of range, precision, time delay of analysis, and implementation costs (adapted from \cite{silva2019})}
  \label{fig:Approaches_comparison_crowd_counting}
\end{figure}

To implement overtourism mitigation actions, crowding information should be made available. Several approaches can be used for crowd detection, such as image capturing, sound capturing, social networks, mobile operator’s data, and wireless spectrum analysis \cite{silva2019}. The latter can be performed using passive or active sniffing methods, characterized by exploring protocol characteristics and small information breaches, such as on Wi-Fi or Bluetooth protocols, extensively used in mobile devices. Figure \ref{fig:Approaches_comparison_crowd_counting} provides a comparison of those approaches for crowd counting in terms of range, precision, time delay of analysis, and implementation costs.

The best option regarding cost, precision, and the near-real-time availability of data required for managing tourism crowding effectively, while complying with privacy rights, is the one based on sensing wireless communication traces, since the vast majority of tourists carry a mobile phone \cite{silva2019,singh-2020}. Earlier approaches relied on counting the number of unique MAC (Media Access Control) addresses in messages emitted by mobile devices. However, due to user privacy concerns, most mobile devices nowadays use MAC address randomization, i.e., the same device exposes different MAC addresses over time, making it more challenging to accurately count the number of devices, thus leading to inaccurate crowd counting.

This chapter describes a low-cost approach to monitoring overtourism. It consists of a crowding sensor that performs real-time detection of trace elements generated by mobile devices from different wireless technologies, namely Wi-Fi and Bluetooth while addressing the MAC address randomization issue when determining the number of mobile devices in the sensors' vicinity, as an improvement over our previous work \cite{silva2019}. Another improvement refers to the provision of multiple communication methods for uploading the crowding information to a cloud server, by using either Wi-Fi or LoRaWAN protocols, thus mitigating network limitations on the installation location, something that has been disregarded on other approaches.

Our sensor is the basis of a Smart Tourism Toolkit (STToolkit) being built in the scope of the \href{https://www.resetting.eu}{RESETTING}\footnote{RESETTING is an acronym for ``Relaunching European smart and SustainablE Tourism models Through digitalization and INnovative technoloGies''} project, funded by the \href{https://single-market-economy.ec.europa.eu/smes/cosme\_en}{European COSME Programme}, to facilitate the transition towards a more sustainable operation of tourism SMEs, and improved quality of the tourism experience. The STToolkit will guide how to build and set up our sensors, either in indoor or outdoor appliances, by including support materials such as an installation manual, video tutorials, setup images, and cost calculators.

Furthermore, this research considers a correlation between the number of mobile devices and the real number of people present in an area. This assumption is especially relevant in touristic scenarios, where our sensors are aimed to be deployed since tourists usually carry their mobile phones to take pictures and record videos during their visits. Therefore, it is assumed that the number of mobile devices in a given area is directly correlated with the number of people in the same area. This is corroborated by \cite{de2016assessing}, where it is shown that mobile phone data is a valuable data source for statistical counting of people.

This chapter is organized as follows: section \ref{sec:RelatedWork} identifies and discusses related work; section \ref{sec:Proposed System architecture} presents the proposed architecture of a typical installation using our sensors; then, in section \ref{sec:Proposed Wi-Fi detection algorithm}, we describe our proposed Wi-Fi detection algorithm which tackles the MAC address randomization issue; on section \ref{sec:Adopted-technologies}, we describe the technologies used in our solution; then, on section \ref{sec:Validation} we present the setup and discuss the results obtained from a field validation; finally, on section \ref{sec:Conclusion}, we draw some conclusions and outline future work.

\section{Related work}
\label{sec:RelatedWork}

\begin{figure}[!htb]
  \centering
  \includegraphics[scale=0.31]{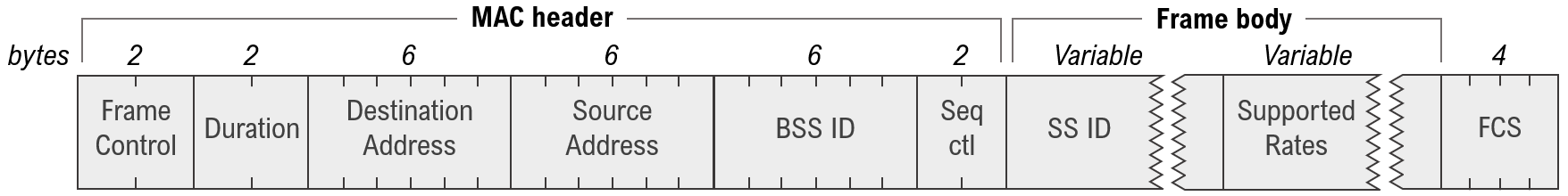}
  \caption{Probe request frame (based on \cite{IEEE802.11})}
  \label{fig:Probe_request_frame}
\end{figure}

Crowd counting by detecting trace elements from mobile devices’ wireless activity can be performed either by the use of passive or active sniffing methods. However, only passive methods, that monitor wireless traffic in a non-intrusive manner, are acceptable, because active methods can cause network and user disruptions, as well as legal infringements. Many passive methods employ probe request capturing, which are messages periodically sent by devices to announce their presence to surrounding APs (Access Points), allowing a fast connection upon reaching a known network. These messages are sent in bursts and are unencrypted, meaning that they can be simply captured using passive sniffing techniques, and contain the device’s MAC address. The probe request frame structure is presented in Figure \ref{fig:Probe_request_frame}.

Earlier detection approaches relied on the device’s real MAC address that was sent in the SA (Source Address) of these frames. In this case, the number of devices was simply equal to the number of different MAC addresses. However, when devices send their real MAC address, they may be easily tracked. To solve this privacy vulnerability, since 2014, manufacturers started to implement MAC address randomization on their devices. It consists of assigning to probe requests randomly generated virtual MAC addresses changing over time. Thus, the real MAC address remains unknown, protecting the user's identity and making it much more difficult to track. Unfortunately, this has led to inaccurate crowd counting and has hampered many solutions adopted until then. Moreover, the MAC address randomization process is dependent on the manufacturer and the operating system of the device, which also makes it a much more complex procedure to circumvent.

\begin{figure}[!htb]
  \centering
  \includegraphics[scale=0.3]{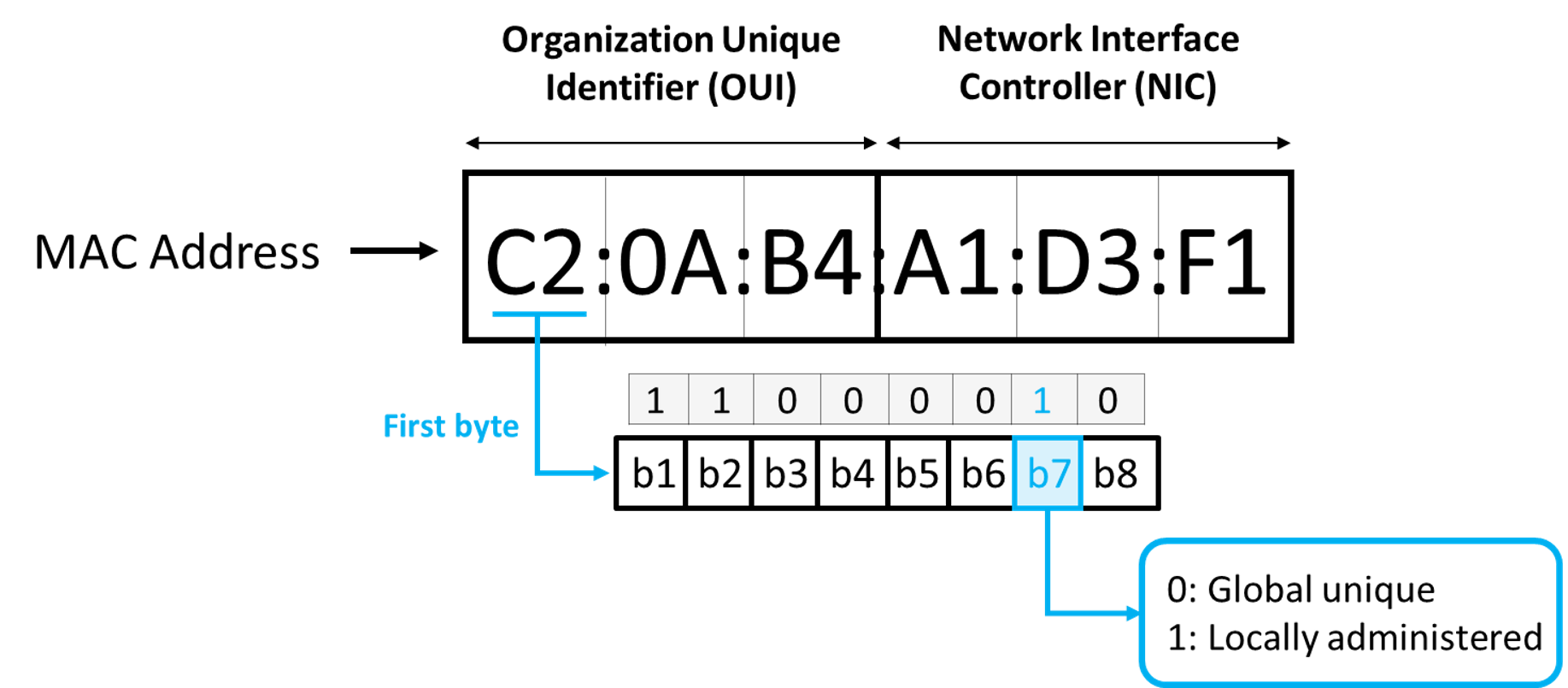}
  \caption{Difference between a real and a virtual MAC address (adapted from \cite{uras-2022})}
  \label{fig:real_MAC_virtual_MAC}
\end{figure}

The difference between a real MAC address (globally unique) and a virtual MAC address (locally administered) is in the \nth{7} bit of the first byte of the MAC address, as shown in Figure \ref{fig:real_MAC_virtual_MAC}. Therefore, we can simply distinguish these two types of MAC addresses by only checking this bit.

The implementation of the MAC address randomization added a level of complexity to uniquely identify devices. Therefore, the research has advanced towards the exploration of other properties and fields of the probe request frames, since the MAC address is no longer a reliable option just by itself for accurate crowd counting. In spite of this, probe request frames still disclose other weaknesses that can be exploited for counting the number of devices. Several strategies have been adopted to mitigate the impact of randomization, as follows:

\begin{itemize}
    \item \textbf{SSIDs\footnote{SSID (Service Set IDentifier) is a sequence of characters that uniquely names a Wi-Fi network} Comparison:} based on comparing the known networks (in terms of SSIDs) to a device, an information that is contained in the probe requests \cite{berenguer-2022}.  
    \\
    \item \textbf{Fingerprinting:} based on generating a unique identifier (fingerprint) from other fields in probe request frames. The contents to generate this fingerprint are usually obtained from IEs (Information Elements) conveyed in the frame body of probe requests. \cite{bravenec-2022, vega-barbas-2021}  
    \\
    \item \textbf{Fingerprinting + Clustering:} this strategy relies not only on fingerprinting from IEs, but also on other properties from these messages, such as the SEQ (SEQuence number), burst size or IFTA (Inter-Frame Time Arrival) from probe request frames. A clustering algorithm considering these properties simultaneously can be applied, where each cluster will represent a unique device. \cite{Cai2021, andrei-2022, he-2022, sospedra-2023, uras-2020, uras-2022}
\end{itemize}

In \cite{berenguer-2022} the PNL (Preferred Network List), which contains the SSIDs from the known networks of a device sent in probe requests, is used for counting the number of devices in a location and distinguishing residents from visitors in the city of Alcoi in Spain. The authors claim an accuracy of 83\% in detection, with some reported overestimations and incongruencies. 

Most approaches rely, however, on applying fingerprinting techniques to uniquely identify mobile devices. The work reported in \cite{vega-barbas-2021} used a network of sensors to estimate the number of persons in a given location based on IEs fingerprinting. The system was tested in public events with a considerable density of people, with a claimed accuracy close to 95\%. Another work described in \cite{bravenec-2022} used the same approach, considering not only the IEs but also the PNL and the recurrence of the same randomized MAC address to generate device fingerprints at a conference in Lloret de Mar, Spain, but precision is not reported.

Some other studies not only considered IEs for fingerprinting but also clustered this information along with other properties or patterns from probe requests. For this purpose, many studies used clustering algorithms that consider a combination of different features from probe requests. In \cite{Cai2021}, not only probe requests, but also beacons and data packets were used to count the number of devices in given locations. This method was tested with a dataset purposefully generated for the scope of this work, reaching an accuracy of 75\%, however, it was not tested in a real crowded scenario. The work reported in \cite{he-2022} considered IEs, SEQ, and the RSSI (Received Signal Strength Indicator) from probe requests with a neural network for estimating crowding levels in a shopping mall in Hong Kong, reaching an accuracy of slightly over 80\%.

The studies reported in \cite{uras-2020, uras-2022} clustered fingerprinting from IEs, the incremental speed of the SEQ, the burst frequency, and the IFTA. The authors first tested the algorithm at the University of Cagliari’s Campus \cite{uras-2020}, achieving an accuracy of about 91\%. A follow-up to this work \cite{uras-2022} tested the algorithm first in a controlled environment, reaching an accuracy of 97\%, and further inside buses in Italy for an Automatic Passenger Counting system, with a precision of 75\%. Another work reported in  \cite{andrei-2022} used the same approach considering the IEs for fingerprinting and also used a clustering algorithm for combining the generated fingerprints with burst sizes and the IFTA in the canteen of the University of Twente, with an accuracy of 90\%. The work described in \cite{sospedra-2023} combined fingerprints from RSSI values of Wi-Fi APs and BLE (Bluetooth Low Energy) beacons with several clustering algorithms variants for indoor positioning, achieving a precision of around 93\%.

Table \ref{table:Primary_studies_MAC_randomization} summarizes the previous approaches for crowd counting, clarifying the adopted strategies to mitigate MAC address randomization and the obtained precision. 

\renewcommand{\arraystretch}{1.45}

\newcolumntype{P}[1]{>{\centering\arraybackslash}p{#1}}
\newcolumntype{M}[1]{>{\centering\arraybackslash}m{#1}}

\definecolor{myLightgray}{rgb}{0.9019,0.9019,0.9019}
\definecolor{myLightLightgray}{rgb}{0.9511,0.9511,0.9511}

\begin{table}[H]
\caption{Approaches for crowd counting, tackling MAC address randomization}
\centering
\begin{tabular}{| M{1.5cm} | M{2cm} | M{3cm} | M{2.5cm} | M{2cm}  |}
\hline
\rowcolor{myLightgray}\textbf{Authors} & \textbf{Packet type capturing} & \textbf{Strategies for MAC address randomization} & \textbf{Real Scenario Appliance} & \textbf{Precision} \\   
\hline
\cite{berenguer-2022} & Probe Requests & SSIDs Comparison & Alcoi (Spain) & 83\% \\ 
\hline
\cite{vega-barbas-2021} & Probe Requests & Fingerprinting & Public events & 90\% \\ 
\hline
\cite{bravenec-2022} & Probe Requests & Fingerprinting & Conference at Lloret del Mar (Spain) & Not available \\ 
\hline
\cite{Cai2021} & \parbox{2cm}{\centering Beacons\\Data packets\\Probe Requests} & Fingerprinting + Clustering &  Not available & 75\% (with simulated data) \\ 
\hline
\cite{he-2022} & Probe Requests & Fingerprinting + Clustering & Shopping mall in Hong Kong & 80\% \\
\hline
\cite{uras-2020} & Probe Requests & Fingerprinting + Clustering & Campus of the Univ. of Cagliari & 91\% \\
\hline
\cite{uras-2022} & Probe Requests & Fingerprinting + Clustering & Buses in Italy & 75\%
 \\
\hline
\cite{andrei-2022} & Probe Requests & Fingerprinting + Clustering & University of Twente campus & 90\% \\
\hline
\cite{sospedra-2023} & Beacons & Fingerprinting + Clustering & Not available & 93\% \\
\hline
\end{tabular}
\label{table:Primary_studies_MAC_randomization}
\end{table} 

\section{Proposed system architecture}
\label{sec:Proposed System architecture}

\begin{figure}[!htb]
  \centering
  \includegraphics[scale=0.55]{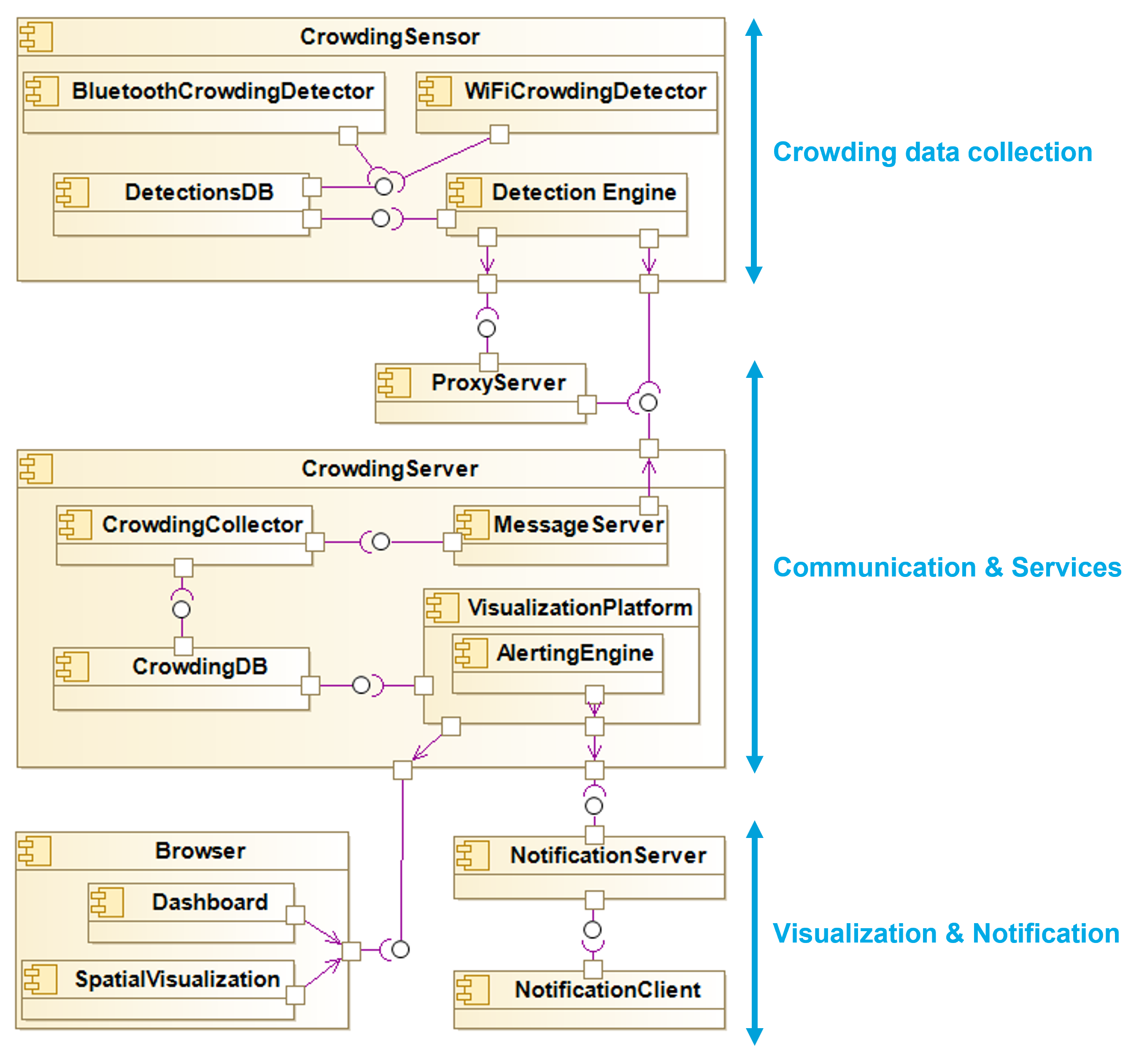}
  \caption{Component diagram of the crowding detection STToolkit}
  \label{fig:Component_diagram}
\end{figure}

The proposed system architecture of the STToolkit is presented in Figure \ref{fig:Component_diagram}, where sensors count the number of devices in their vicinity and periodically report the crowding information to a cloud server. The latter also has other components for making downlink communication transparent and providing uplink services for rendering the crowding information and creation of notification policies.

\subsection{Crowding data collection}
\label{subsec:Crowding data collection}
Each crowding sensor includes a detector responsible for passively capturing mobile devices' trace elements for each wireless technology (Wi-Fi and Bluetooth) in the sensor vicinity, an anonymized local database, where all gathered information is stored, and a detection engine, responsible for counting the number of devices by analyzing the information contained in the local database and reporting the crowding information to the cloud server. Each sensor can perform only Wi-Fi or Bluetooth detection, or both simultaneously, and can quickly switch between the technologies to be used for detection.

In this edge computing approach, data collection and crowding level measurement generation are performed locally in each sensor, so that only the number of devices detected is sent to the cloud server. As so, the information to be passed is minimal, not requiring a high sampling rate for data transmission, and also protecting nodes from outside threats, since the communication line prevents the majority of attack types. Furthermore, limiting data exchange not only reduces communication costs, but also eases protection complexity for the node, and makes it easier to guarantee user privacy. Also regarding user privacy, all gathered data is anonymized before being stored in the local database.

\subsection{Communication and services}
\label{subsec:Communication and Services}
To better address installation location requirements and connectivity limitations, a flexible deployment regarding uplink technologies has been considered. Data can be uploaded to the cloud server by using a variety of communication protocols, such as Wi-Fi or LoRaWAN (Long Range Wide Area Network). 

If Wi-Fi is available on site, data can be uploaded directly to the \emph{Message Server} via the MQTT (Message Queuing Telemetry Transport) protocol, a lightweight method of carrying out messaging, using a publish/subscribe model, widely used for IoT (Internet of Things) applications. This option can be applied straightforwardly in indoor tourism scenarios, for instance, in a museum, which generally provides a Wi-Fi network to visitors. 

In outdoor scenarios, such as public parks or city squares, where overtourism situations can also arise, Wi-Fi coverage may not be available. Since sensors must upload crowding information, other approaches rely on mobile operators' communication, which may be an expensive option, usually with monthly fees depending on the number of sensors used, each using a SIM card.

\begin{figure}[!htb]
  \centering
  \includegraphics[scale=0.50]{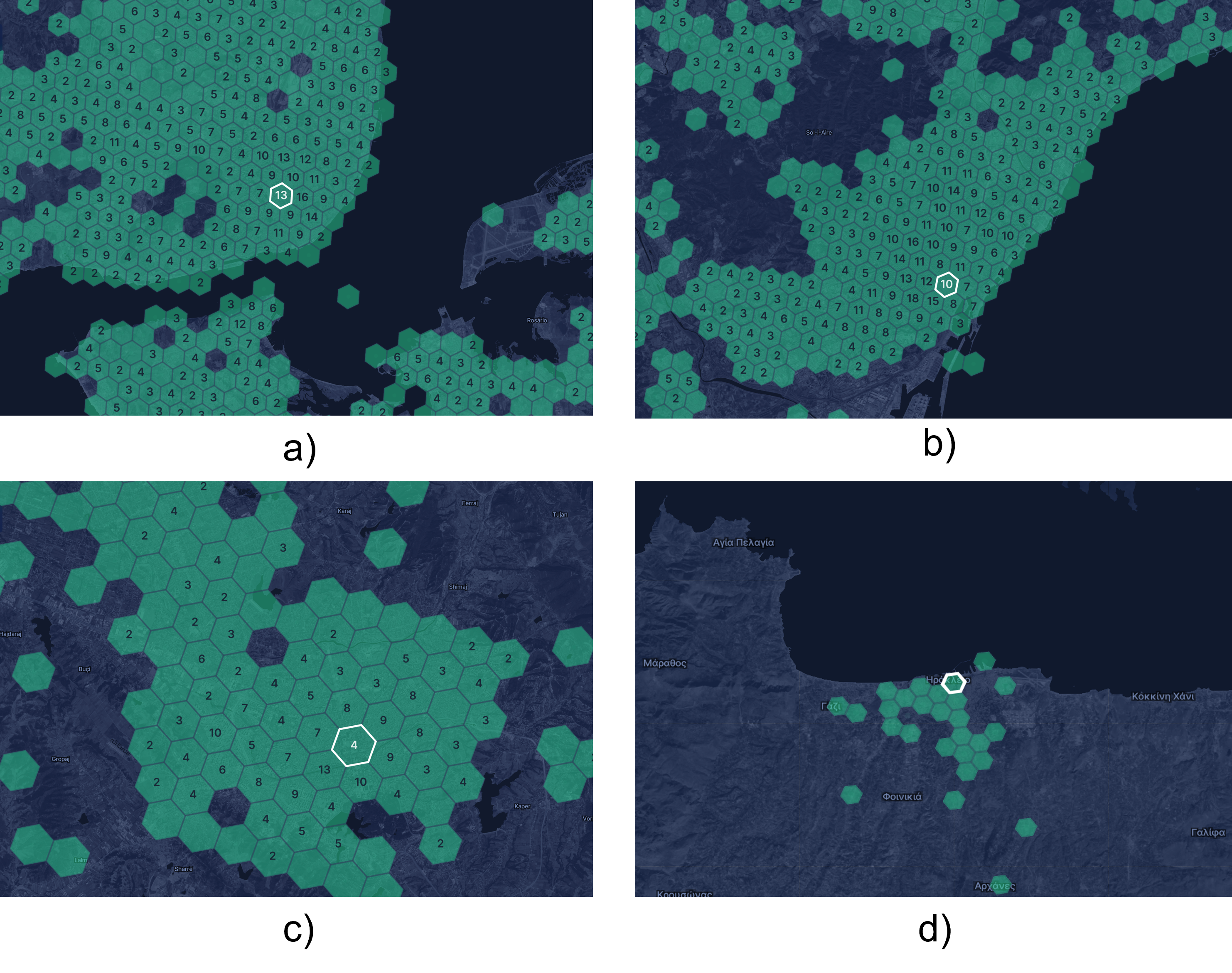}
  \caption{Helium network coverage in: a) Lisbon, b) Barcelona, c) Tirana, d) Heraklion}
  \label{fig:Helium_coverage}
\end{figure}

To mitigate this problem we offer the option for uploading data via LoRaWAN, a standard of the International Telecommunication Union that provides a low-cost and scalable alternative that is feasible for our application, since sensors only communicate a small amount of data, i.e., the number of detected devices. For this, sensors must be equipped with a LoRa board and corresponding antenna to communicate the crowding information to a LoRaWAN gateway that, in turn, will route the information to the cloud server via the MQTT protocol. Regarding coverage, there are a few LoRa networks, designed for IoT appliances, that can be used for uploading data, like \href{https://www.thethingsnetwork.org/}{The Things Network} open collaborative network or the, also crowdsourced, \href{https://www.helium.com/}{Helium} network, a decentralized wireless infrastructure supported by blockchain. The Helium network adopted for this solution is the fastest growing IoT network with LoRaWAN compatibility that provides a large coverage in many countries in Europe, such as those involved in the RESETTING project. Figure \ref{fig:Helium_coverage} shows the \href{https://explorer.helium.com}{Helium network coverage} in cities where our STToolkit may be deployed in the context of the RESETTING project, such as Lisbon, Barcelona, Tirana, and Heraklion, the capital of the Greek island of Crete.

Regarding the Helium network, an SME can choose between two alternatives for uploading crowding information: using Helium with a 3rd party LoRaWAN service provider, such as \href{https://www.helium-iot.eu}{Helium-IoT}, or using Helium with a private LoRaWAN server. 

As shown in Figure \ref{fig:Component_diagram}, the \emph{Message Server} is the only entrance point for all messages in the cloud server, independently of the communication protocol used for uploading the crowding information. This provides transparency since all messages are received in the cloud server via the MQTT protocol independently of the communication technology adopted for uploading the crowding information. 

Furthermore, in areas with low or no Wi-Fi or Helium network coverage, it is also possible to acquire equipment for that purpose, such as a Wi-Fi mesh system, that will allow expanding the Wi-Fi network coverage, or a Helium hotspot, for grating Helium network coverage for uploading data via the LoRaWAN protocol.

\subsection{Visualization and notifications}
\label{subsec:Visualization and Notifications}
The cloud server also has several components to make downlink communication transparent and provide several uplink services that can be used by Smart Tourism Tools to understand the crowding levels in areas where each sensor is placed, with a clear and simple perspective. Possible crowding services are:

\begin{itemize}
    \item Rendering of temporal information, as seen in Figure \ref{fig:Comparing-crowding}.
    \item Rendering of geographic information, as seen in Figure \ref{fig:Crowding-hotspots-at-Iscte}.
    \item Notification policies, e.g., when crowding threshold levels are reached.
    \item Raw data for custom-made integrations, e.g., spatial visualization using a BIM (Building Information Model), as seen in Figure \ref{fig:ISCTE_BIM_model}.
\end{itemize}

\section{Proposed Wi-Fi detection algorithm}
\label{sec:Proposed Wi-Fi detection algorithm}

MAC address randomization performed by mobile device manufacturers, due to user privacy concerns, has made the identification of a mobile device a much more difficult task and, consequently, more difficult to accurately perform device counting. Therefore, an algorithm was developed for the detection of mobile devices through Wi-Fi, tackling the MAC address randomization issue using a fingerprinting technique, presented in Figure \ref{fig:proposed_Wi-Fi_detection_algorithm}. The explanation of each step of the proposed algorithm is presented below. A similar algorithm is also envisaged for Bluetooth detection since the randomization problem is also pertinent to this technology.

\begin{figure}[!htb]
  \centering
  \hspace*{-0.2cm}
  \includegraphics[scale=0.36]{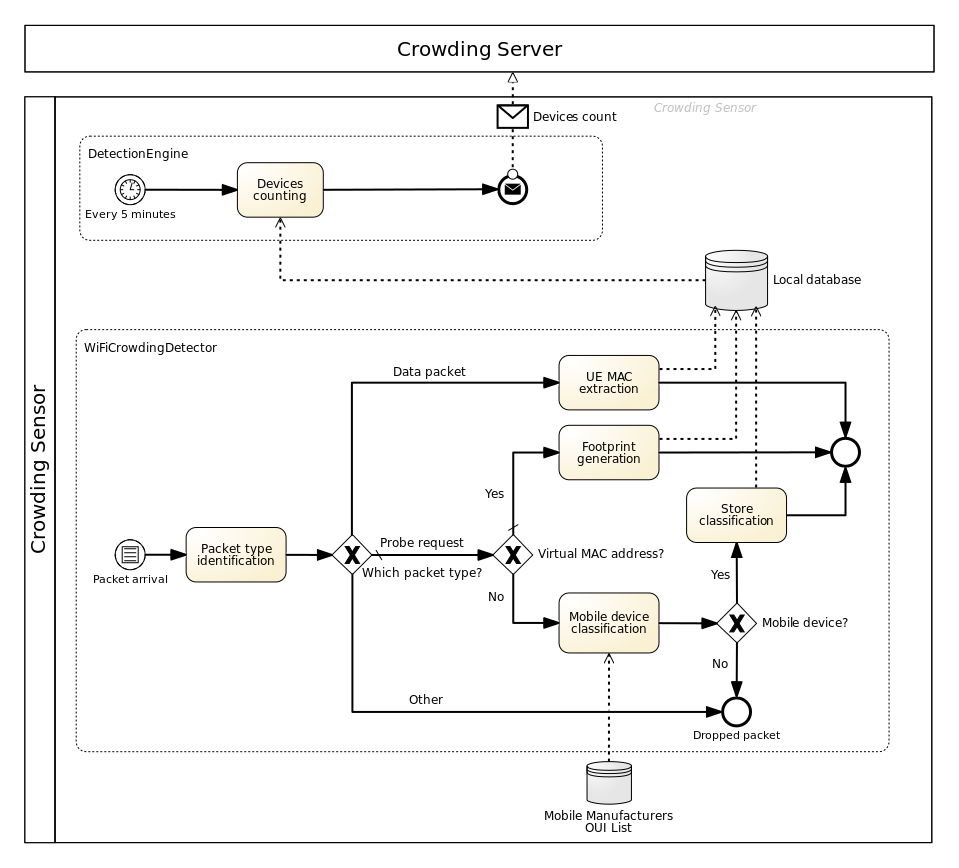}
  \caption{Proposed Wi-Fi detection algorithm}
  \label{fig:proposed_Wi-Fi_detection_algorithm}
\end{figure}

The \emph{WiFiCrowdingDetector}, seen in Figure \ref{fig:Component_diagram}, is responsible for processing each Wi-Fi packet captured by the sensor. The first operation performed is a packet type identification (data packets, probe requests, or other type). Data packets will be used for counting the number of devices connected to an AP, and probe requests for counting the number of devices not connected to any AP. The packet type can be obtained by checking the Frame Control field, presented in Figure \ref{fig:Probe_request_frame}. Packets that are not either data packets or probe requests will be immediately dropped since they are not relevant for device counting.

Regarding data packets, only the UE (User Equipment) part of the MAC address needs to be accounted for. First, it is necessary to locate it in the frame, which is performed by checking the DS (Device Status) information in the frame header, since the UE MAC address position may vary according to the direction of the frame. These MAC addresses can be directly counted as single devices because when a device is connected to an AP, the MAC address is kept constant throughout the connection and, therefore, will not change randomly. For this reason, after the UE MAC address extraction, it is directly stored in the sensor’s anonymized local database.

Regarding probe requests, the first operation performed is aimed at distinguishing its Source Address between a real and a virtual MAC address. This is performed by checking the \nth{7} less significant bit of the \nth{1} octet of the MAC address, as already shown in Figure \ref{fig:real_MAC_virtual_MAC}. To follow the trace of a real MAC address, a device classification is applied, aimed at only counting MAC addresses that belong to mobile devices. This is done by checking the address's OUI (Organizational Unique Identifier)\footnote{OUI is a part of the MAC address identifying the network adapter vendor}: if the OUI matches one of the known mobile manufacturers, obtained from the \href{https://gitlab.com/wireshark/wireshark/-/raw/master/manuf}{Wireshark manufacturer database}, the MAC address should be considered as a mobile device and it must be counted, and stored in the local database; otherwise, the MAC address is not considered as a mobile device and is discarded. 

To follow the trace of a virtual MAC address, a fingerprinting technique must be performed to uniquely identify devices that use MAC address randomization. For this, the IEs contained in the frame body of the probe request are analyzed. For each IE, the entirety of its information is considered, including the IE ID, Length, and Value bytes. For those IEs with substantially varying values across probes emitted from the same device (e.g., DS Parameter Set), only the bytes of IE ID and Length are analyzed. Table \ref{table:Information_Elements_used_fingerprinting} shows the IEs used for the fingerprinting technique. After analyzing all IEs, a hash function is applied to all its contents. As a result, a 64-bit footprint is generated for each probe request and stored in the local anonymized database. Then, all the devices that are trying to connect to a Wi-Fi network, by sending probe requests to discover available networks in proximity with a virtual MAC address, are uniquely identified by the footprint. As so, each footprint should be counted as one mobile device using MAC Address randomization that is trying to connect to a Wi-Fi network. To avoid counting the same device twice, if the same MAC address is captured in both data packets and probe requests, it is only accounted for once.

\renewcommand{\arraystretch}{1.4}

\begin{table}[H]
\centering
\caption{Probe Request's Information Elements used to create device fingerprint}
\begin{tabular}{ p{3.2cm} >{\centering\arraybackslash}p{1cm} >{\centering\arraybackslash}p{1cm} p{5.3cm} }
\hline
\multicolumn{1}{c}{\textbf{Information Element}} & \multicolumn{1}{c}{\textbf{IE ID}} &\multicolumn{1}{c}{\textbf{IE Length}} &\multicolumn{1}{c}{\textbf{Description}}\\
\hline

Supported Rates & 1 & <8 & Data transfer rates supported by the device. \\ 
Extended Supported Rates & 50 & <256 & Other bit rates supported by the device. \\ 
DS Parameter Set & 3 & 1 & Device's channel setting when sending a probe request. \\ 
HT Capabilities & 45 & 26 & Compatibility with the 802.11n standard. \\
VHT Capabilities & 191 & 12 & Compatibility with the 802.11ac standard. \\
Extended Capabilities & 127 & <256 & Other device capabilities. \\
RM Enabled Capabilities & 70 & 5 & Information for measuring radio resources. \\
Interworking & 107 & <9 & Interworking service capabilities
of the client. \\ 
Vendor Specific & 221 & <256 & Vendor Specific information (e.g., device manufacturer) \\ 
\hline

\end{tabular}
\label{table:Information_Elements_used_fingerprinting}
\end{table}

Then, the \emph{Detection Engine} will periodically count the number of devices detected within a sliding window of X minutes, i.e., the number of devices detected in the last X minutes, and upload that information to the cloud server. Both the sliding window period and data sampling rate can be independently and easily configured by the user. Since we intend to provide real-time or near-real-time data availability, the data sampling rate of our sensors needs to comply with this requirement. As so, we have chosen a 5-minute period for the data sampling rate, as it is a sufficient time period for providing near-real-time data availability. Also, the same 5-minute period was chosen for the sliding window, so that each crowding measurement could comprise all detected devices within each sliding window. 

The number of devices detected is the sum of (i) the number of devices connected to an AP, obtained from the UE MAC addresses captured in data packets, plus (ii) the number of different devices not connected to any AP, obtained from the MAC addresses from probe requests with real MAC addresses, plus (iii) the number of different footprints from probe requests with virtual MAC addresses.

\section{Adopted technologies}
\label{sec:Adopted-technologies}

\begin{figure}[!htb]
  \vspace{1em}
  \centering
  \includegraphics[scale=0.35]{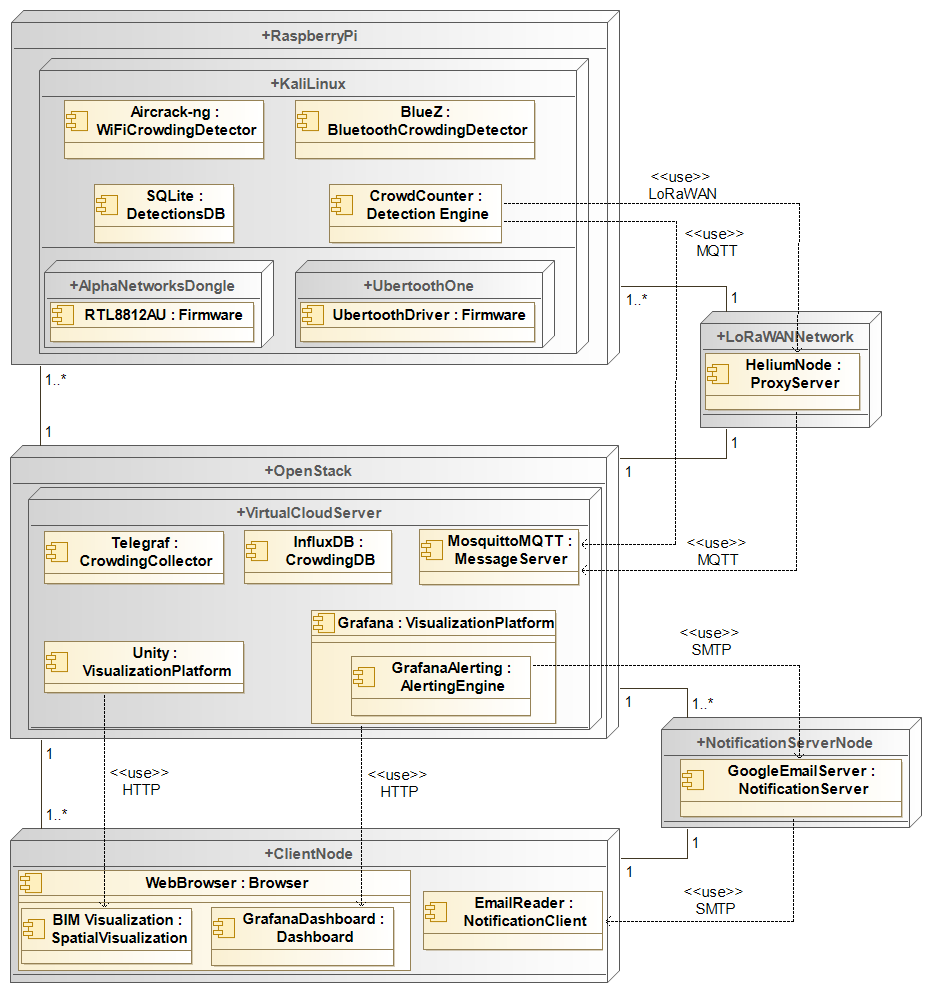}
  \caption{Deployment diagram for the crowding detection STToolkit}
  \label{fig:Deployment_diagram}
\end{figure}

The developed STToolkit uses a variety of open-source software technologies, installed in off-the-shelf hardware available at affordable costs. Figure \ref{fig:Deployment_diagram} presents the UML deployment diagram proposed for the STToolkit concerning all technologies adopted in our solution.

For the operating system of our sensors, we have opted for a \href{https://www.kali.org}{Kali Linux} distribution, which comes with a large number of preinstalled network tools that can be easily used for detecting devices in different technologies. For the local database, a \href{https://sqlite.org}{SQLite} database was chosen for storing all gathered data, which requires low memory usage, while meeting all other requirements. For data anonymization, the sensors use the \href{https://github.com/erthink/t1ha}{t1ha} library that provides several terraced and fast hash functions. In particular, we have opted for the t1ha0 hash function, as it is one of the fastest available at the library.

For performing Wi-Fi detection, the required hardware is a Wi-Fi card that supports monitor mode, which allows the board to capture all network traffic in its proximity. We have chosen the \href{https://alfa-network.eu/alfa-awus036ac}{Alfa Network AWU036AC} board for our sensor, which provides high performance at a low cost, having two antennas for dual-band detection (2.4 GHz and 5 GHz) without interfering with Bluetooth devices. As for the sniffing software, we have chosen the \href{https://www.aircrack-ng.org}{Aircrack-ng} tool, an open-source software with several different applications for detecting devices. In particular, we use \emph{airmon-ng} for enabling the monitor mode in the Wi-Fi board, and \emph{airodump-ng} for capturing raw Wi-Fi frames. For performing Bluetooth detection, we have selected the \href{https://greatscottgadgets.com/ubertoothone}{Ubertooth-One} board and corresponding \href{https://www.kali.org/tools/bluez/}{BlueZ} package that contains tools and frameworks for Bluetooth usage in Linux. 

For receiving all messages, our cloud server uses the \href{https://mosquitto.org}{Mosquitto MQTT}, a lightweight message broker that implements the MQTT protocol.

For the data ingestion of all measurements sent by sensors, a database is necessary. This database has to be lightweight, capable of querying data rapidly from timestamps, and also capable of providing support for data visualization platforms to observe the results in real-time. That is why we chose \href{https://www.influxdata.com}{InfluxDB}, a time-series database focused on IoT applications, for our \emph{CrowdingDB} component. For the \emph{CrowdingCollector}, responsible for pushing all messages received in the \emph{Message Server} via MQTT protocol to the \emph{CrowdingDB} in the appropriate format, we chose \href{https://www.influxdata.com/time-series-platform/telegraf}{Telegraf}, an open-source plugin-driven server agent for collecting and reporting metrics from devices.

Finally, for data visualization, we chose \href{https://grafana.com}{Grafana}, an open-source analytics and monitoring tool compatible with several databases, including \emph{InfluxDB}. This framework can be used for creating custom dashboards with graphs and panels for viewing, with different spatio-temporal levels of granularity, the crowding information. Additionally, \emph{Grafana} can be used for creating notification policies, allowing users to receive alerts according to the crowding levels via a diversity of contact points.

\begin{figure}[!ht]
  \centering
  \includegraphics[scale=0.62]{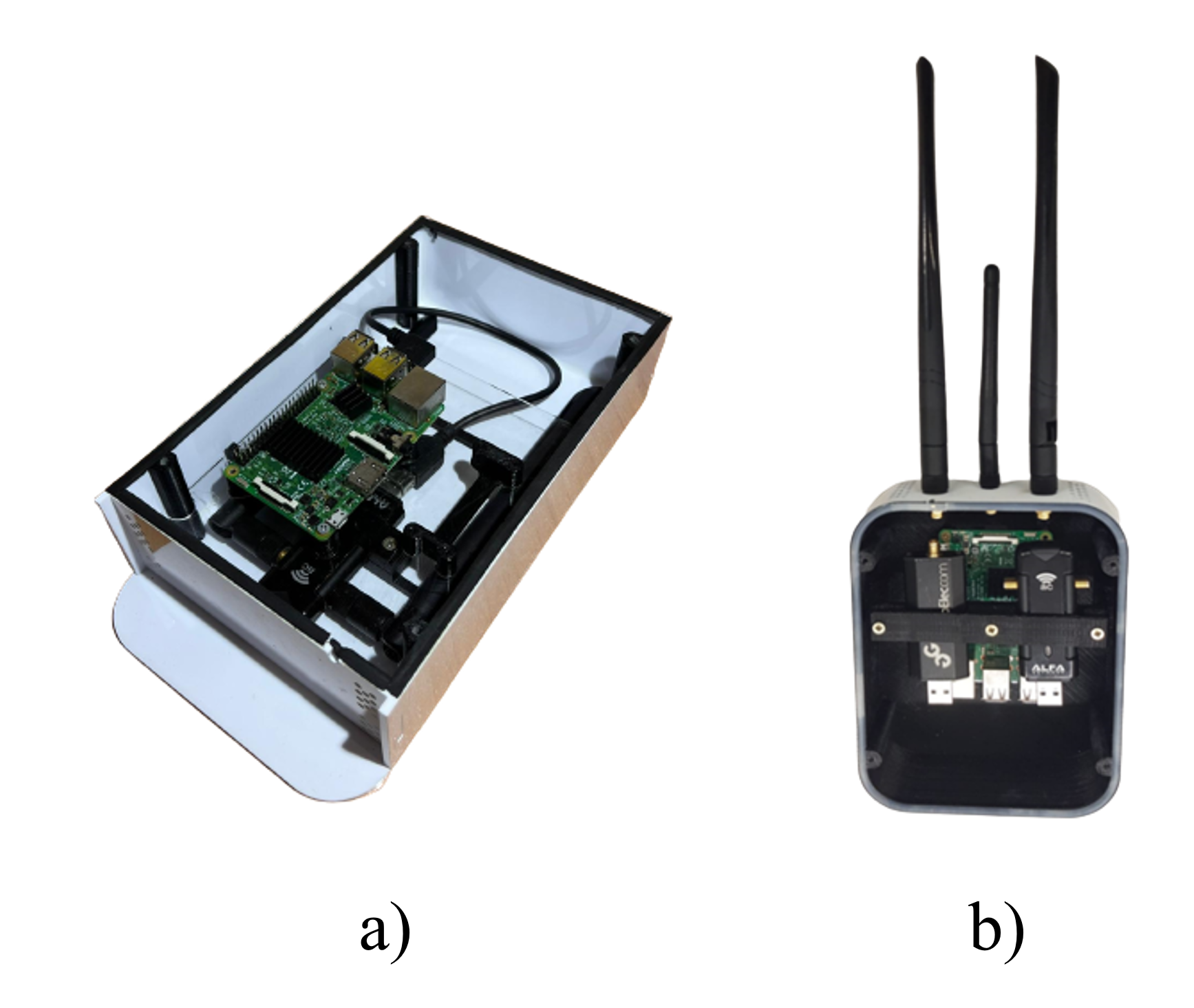}
  \caption{Sensor cases: a) large version with no exposed antennas; b) small version with exposed antennas}
  \label{fig:Sensor-cases}
\end{figure}

A prototype was developed whose hardware components and respective functions are illustrated in Table \ref{table:Hardware_components}. The prototype uses custom-designed cases adapted to deployment locations, either with exposed antennas or not, as shown in Figure \ref{fig:Sensor-cases}. These prototype versions have been deployed at several locations at our university campus to test the operation and performance of the STToolkit, which is further described in the next section.

\renewcommand{\arraystretch}{1.5}

\newcolumntype{M}[1]{>{\centering\arraybackslash}m{#1}}

\begin{table}[H]
\centering
\caption{Prototype hardware components and respective functions}
\begin{tabular}{| M{4.5cm} | M{5.5cm} | }
\hline
\rowcolor{myLightgray}\multicolumn{1}{|c|}{\textbf{Component}} & \multicolumn{1}{|c|}{\textbf{Function}}\\
\hline
Raspberry Pi 3/4 & Coordinate and Process \\
\hline
Alfa Network AWUS036AC & Wi-Fi detection \\
\hline
Ubertooth-One & Bluetooth detection \\
\hline
Raspberry Pi IoT LoRa pHAT & Upload via LoRaWAN (if necessary)\\
\hline
\end{tabular}
\label{table:Hardware_components}
\end{table}

Furthermore, as our sensor’s processing unit is a single computer board, namely, a Raspberry Pi, there are multiple options for powering our sensor, either directly from a battery, or even via USB ports, or Power over Ethernet (PoE), even though the most straightforward and convenient alternative should be to directly connect it to a mains power supply through a transformer. 

\section{Field validation and discussion}
\label{sec:Validation}

The prototype described in the preceding section was designed and implemented to withstand all the scenarios where the sensors may be deployed.

To test and validate the STToolkit architecture, sensors have been placed at several spots across Iscte's campus, and crowding information has been collected since September 2022. 

The sensors were deployed both indoors and outdoors, in places with different crowding patterns, such as areas with a large pedestrian flow, internal and external passages between buildings, and places for prolonged stays, such as a large study hall and the university library.

This field experiment has been conducted with the sole purpose of assessing the perception of the crowding phenomena in the university campus, rather than the accuracy regarding the real number of people at each location. It focused on perceiving crowding patterns and tendencies, such as time breaks between classes, lunch periods, and highly populated events. The aim was to assess how sensors could perceive relative variations throughout the days across the several locations of the campus, and how quickly the sensors were able to detect them.

The accuracy was addressed in other contexts in a more controlled environment \cite{santos2023}, where the detections from sensors were compared with the real number of people, obtained through direct observation during a public event, to assess the effectiveness of the solution.

The crowding data has been used for visualization, using a variety of temporal dashboards, and maps that highlight the geographic distribution of crowding at each location where the sensors have been deployed. Data has also been used for spatial visualization in the form of heatmaps and also, for a more realistic view, using avatars on top of Iscte's BIM (Building Information Model). 

\begin{figure}[!htb]
  \vspace{0em}
  \centering
  \includegraphics[scale=0.32]{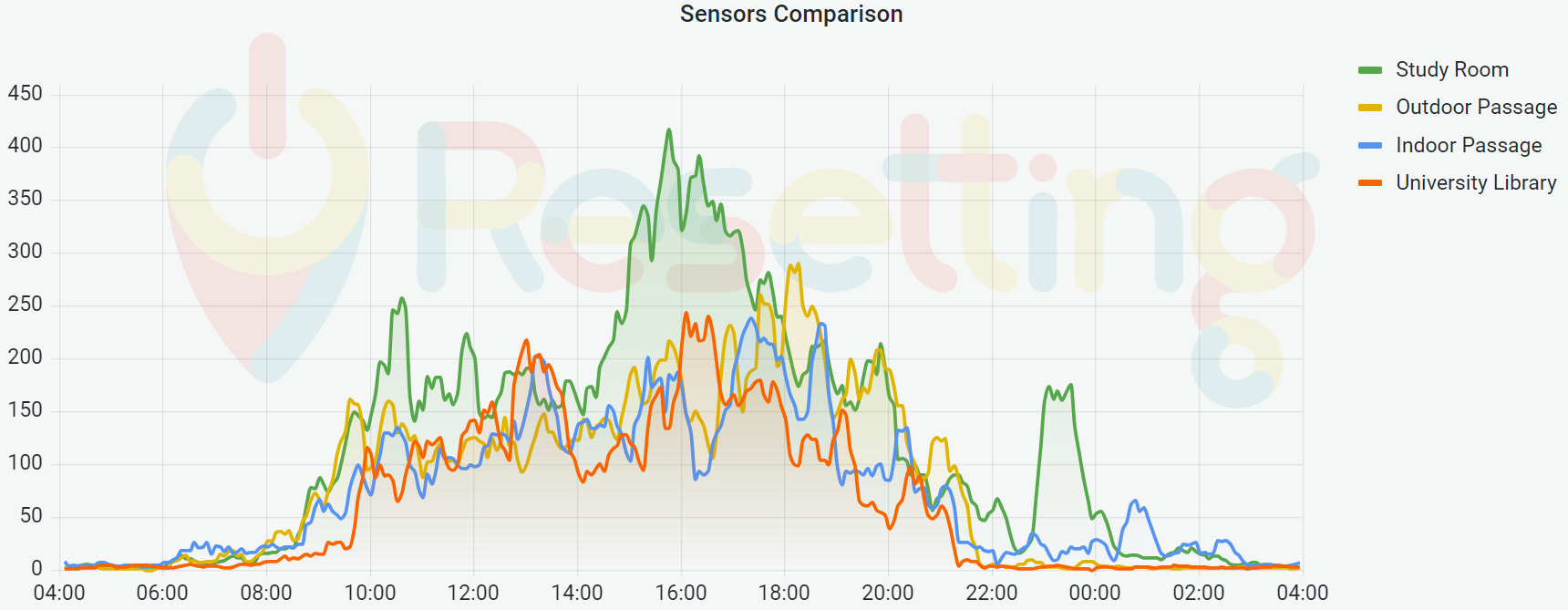}
  \caption{Comparing crowding at Iscte's campus}
  \label{fig:Comparing-crowding}
\end{figure}

Dashboards allow users to select time ranges for crowding data temporal visualization, to perceive people’s concentration and flows during specified periods, and to identify highly populated events. Figure \ref{fig:Comparing-crowding} shows a comparison of crowding levels during a normal day of classes at Iscte’s campus, at the selected spots where sensors have been deployed. 

\begin{figure}[!htb]
  \centering
  \includegraphics[scale=0.40]{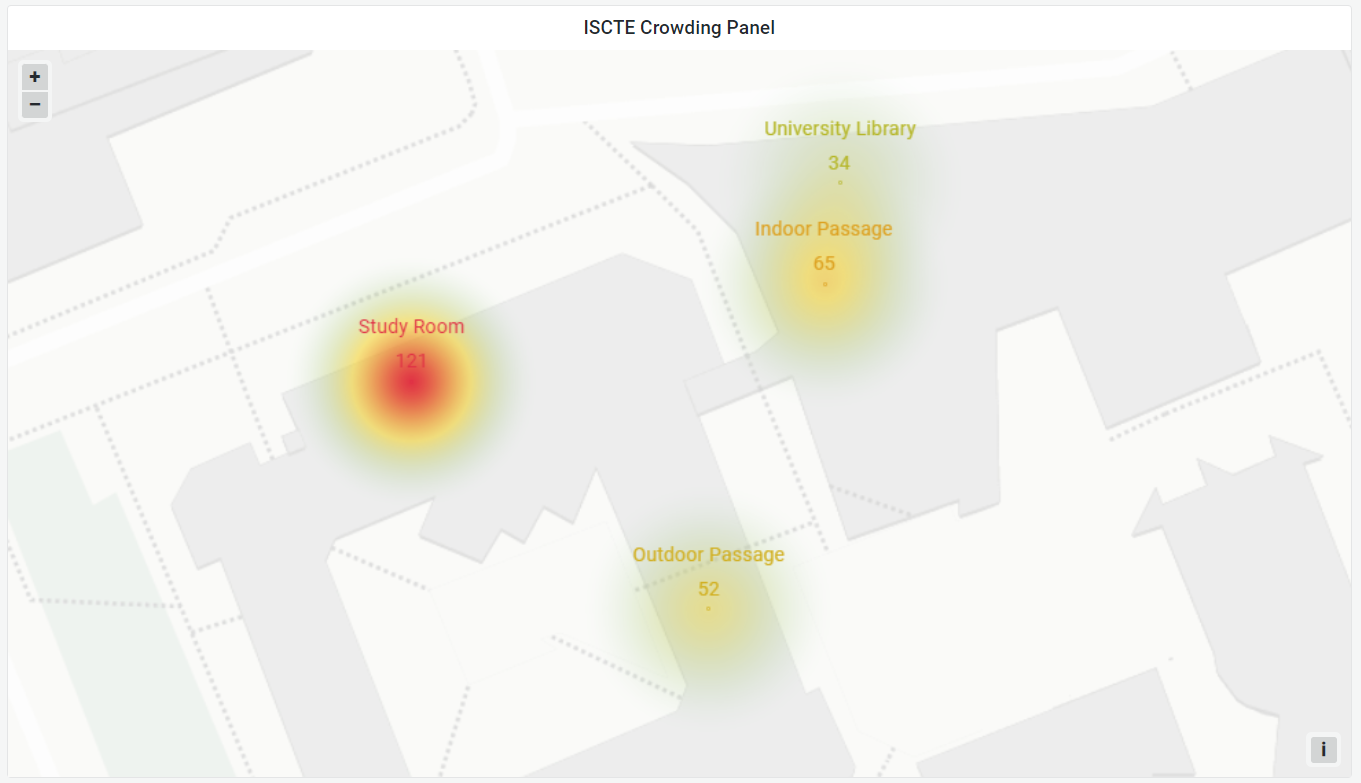}
  \caption{Crowding hotspots at Iscte's campus}
  \label{fig:Crowding-hotspots-at-Iscte}
\end{figure}

In addition to the temporal rendering of the information, data has also been used for spatial visualization in the form of heatmaps, to grant users a better perception of people distribution at several locations where the sensors are deployed. This can be seen in Figure \ref{fig:Crowding-hotspots-at-Iscte}, where it is possible to perceive the crowding hotspots from our sensors deployed at Iscte's campus at a given time.

\begin{figure}[!htb]
  \centering
  \includegraphics[scale=0.56]{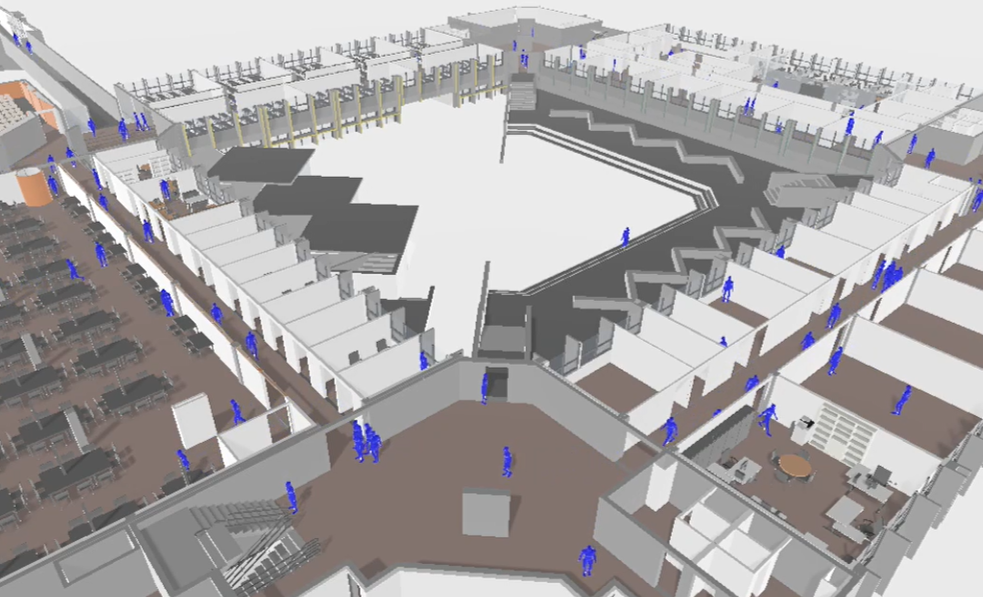}
  \caption{Crowding visualization based on avatars upon the campus BIM}
  \label{fig:ISCTE_BIM_model}
\end{figure}

Moreover, raw crowding information can also be easily used by third-party integrations. To validate this, we built a walking avatar animation upon Iscte's BIM, to achieve a more realistic perception of space occupancy, as shown in Figure \ref{fig:ISCTE_BIM_model}, for one of the campus buildings. There, the number of detected devices, obtained in real-time from sensors, determines the number of ingress and egress avatars in their areas of detection. This last experience was performed during the \href{https://sites.google.com/iscte-iul.pt/resetting-project/home/posters-workshop}{International Posters \& Demos Workshop on Smart Tourism} held by the RESETTING project at Iscte in January 2023.

Furthermore, it is also possible to create notification policies, where alerts can be triggered if predetermined crowding thresholds are exceeded, by using several contact points such as e-mail, \href{https://web.telegram.org/}{Telegram}, \href{https://chat.google.com/}{Google Chat}, \href{https://teams.microsoft.com/}{Microsoft Teams}, \href{https://slack.com}{Slack}, or \href{https://www.pagerduty.com}{PaperDuty}, enabling users to make just-in-time decisions facing overtourism situations. These alerts can be easily configurable by using the \emph{Grafana} tool, also used for spatio-temporal visualization of crowding information.

\section{Conclusions and future work}
\label{sec:Conclusion}

Overtourism deteriorates the visiting experience of tourists, the quality of life of residents, as well as the environment. By monitoring it, tourism professionals can identify areas of concern and put measures in place to lessen its negative effects, encouraging better tourism practices to ensure that tourism benefits both tourists and locals while preserving natural and heritage resources.

For monitoring overtourism, a low-cost approach based on mobile devices' activity has been developed. The sensors, equipped with off-the-shelf hardware available at affordable costs, perform real-time detection of trace elements of mobile devices' wireless activity, mitigating MAC address randomization, and crowding values are put together in a cloud server. Alternative communication channels for uploading the crowding information, namely via Wi-Fi or LoRaWAN protocols, allow for addressing local connectivity limitations at the installation location of sensors. In addition, scalability is provided by maintaining the hardware costs low, by using open-source software, and by the simplicity of the installation and configuration of each sensor. Regarding the RESETTING project, an SME must choose the option that best fits its needs and requirements for implementing its system. To help with this purpose, the STToolkit will include a sensor deployment calculator for SMEs to estimate the most cost-effective uplink alternative according to the installation location of each sensor. 

The crowding information can then be analyzed by destination managers to understand the crowding levels in areas where each sensor is placed in a clear and simple perspective, either by dashboards for temporal or spatial visualization of crowding information or using the raw data for custom-made integrations. Furthermore, notification policies can be created when overtourism situations occur, opening the possibility of implementing just-in-time mitigation actions required by the nature of these circumstances, as they may be sudden and unpredictable.

Preliminary tests have been conducted for this solution. A prototype version of the crowding sensor was deployed at several spots on Iscte's campus, in typical usage scenarios with a high flow and/or extended presence of people, such as the university library, a large study hall, and two passageways. Crowding information has been collected and used to monitor people’s flow and detect high-crowding events on campus.

In the short term, this STToolkit will be deployed at the \href{https://www.parquesdesintra.pt/pt/parques-monumentos/parque-e-palacio-nacional-da-pena/}{Pena Palace}, one of the most iconic tourism sites in Portugal, surrounded by a large walkable park, flagellated by overtourism all year round. The objectives will be promoting alternative routes for tourists within the park, limiting their number in sensitive areas, and making the tourism offer in this area more sustainable. Our detection approach will then contribute to reducing the overwhelming feeling of pressure in critical hotspots, thus leading to a greater visiting experience for tourists who visit this attractive tourist site.

Furthermore, a second prototype version of sensors is also envisaged. The latter will include new boards with greater performance, new antennas with higher gains for larger detection ranges, directional antennas for performing detection in specific areas, custom-designed heatsinks for the processing units to achieve the best possible performance, as well as new custom-designed cases.

Further details on the sensors, including demos of setting up and configuring the edge nodes and the cloud server can be found online at the \href{https://sites.google.com/iscte-iul.pt/resetting-project/home/smart-tourism-toolkits/Crowd_Monitoring_STToolkit}{RESETTING@Iscte site}.

\begin{acknowledgement}
This work has been developed in the scope of the RESETTING project, funded by the European COSME Programme (EISMEA), under grant agreement COS-TOURINN 101038190. The cloud-based infrastructure (computing and storage) used was provided by the INCD, Funded by FCT and FEDER under project 01/SAICT/2016 nº022153. The current work has also been supported by Fundação para a Ciência e Tecnologia (FCT)/Ministério da Ciência, Tecnologia e Ensino Superior (MCTES) through national funds and, when applicable, co-funded by European Union (EU) funds under the project UIDB/EEA/50008/2020.
\end{acknowledgement}

\bibliographystyle{styles/spmpsci}

\bibliography{references}

\end{document}